\documentclass{article}
\usepackage{spconf,amsmath,graphicx,amsthm, mathtools}
\usepackage{multirow}
\usepackage{epstopdf, algorithmicx, algorithm}
\usepackage{epsfig,cite,amssymb,hyphenat}
\usepackage{subfigure}
\usepackage{multirow,breqn,bm,cleveref}
\usepackage{url,dsfont,algpseudocode}
\newtheorem*{lemma*}{}

\newtheoremstyle{mystyle}
  {3pt}{3pt}{\itshape}{}{\bfseries}{}{.5em}
  {\thmname{#1}\thmnumber{ #2}\thmnote{. #3}}
\theoremstyle{mystyle}
\newtheorem*{thm*}{}

\DeclarePairedDelimiter\floor{\lfloor}{\rfloor}

\newcommand{\algparbox}[1]{\parbox[t]{\dimexpr\linewidth-\algorithmicindent}{#1\strut}}
\newcommand{\showfig}[1]{}

\title{Protecting Genomic Privacy by a Sequence-Similarity Based Obfuscation Method}
\name{Shibiao Wan$^{1}$, Man-Wai Mak$^{2}$ and Sun-Yuan Kung$^{1}$\thanks{This work was in part supported by the Brandeis
Program of the Defense Advanced Research Project Agency (DARPA) and
Space and Naval Warfare System Center Pacific (SSC Pacific) under Contract
No. 66001-15-C-4068, and The RGC of Hong Kong SAR, Grant No. PolyU 152068/15E.}}
\address{
$^{1}$Dept. of Electrical Engineering, Princeton University, New Jersey, USA\\
$^{2}$ Dept. of Electronic and Information Engineering, The Hong Kong Polytechnic University, Hong Kong SAR}

\begin{document}
\ninept
\maketitle
\vspace{-0.2cm}
\begin{abstract}

In the post-genomic era, large-scale personal DNA sequences are produced and collected for genetic medical diagnoses and new drug discovery, which, however, simultaneously poses serious challenges to the protection of personal genomic privacy. Existing genomic privacy-protection methods are either time-consuming or with low accuracy. To tackle these problems, this paper proposes a sequence similarity-based obfuscation method, namely IterMegaBLAST, for fast and reliable protection of personal genomic privacy. Specifically, given a randomly selected sequence from a dataset of DNA sequences, we first use MegaBLAST to find its most similar sequence from the dataset. These two aligned sequences form a cluster, for which an obfuscated sequence was generated via a DNA generalization lattice scheme. These procedures are iteratively performed until all of the sequences in the dataset are clustered and their obfuscated sequences are generated. Experimental results on two benchmark datasets demonstrate that under the same degree of anonymity, IterMegaBLAST significantly outperforms existing state-of-the-art approaches in terms of both utility accuracy and time complexity.

\end{abstract}
\vspace{-0.2cm}
\section{Introduction}\label{sec:intro}
\vspace{-0.2cm}

Recent decades have witnessed the widespread applications of genomic high-throughput technologies in personalized healthcare \cite{Chute13}, with which large-scale personal genomic data are produced and collected for genetic medical diagnoses and new drug discovery. Moreover, individuals become more willing to share their genomic data on some health-related websites (e.g., OpenSNP\footnote{https://opensnp.org/}) to learn their predispositions to genetic diseases and their ancestries \cite{Humbert13}. These cases, however, simultaneously pose serious challenges to the protection of personal genomic privacy. Actually, the genomic information of an individual can be as personally indicative as his/her fingerprint, if not more revealing \cite{leonard72}. The genomic information is highly at risk of being abused to affect employment, insurance status, etc \cite{Clayton03}. Due to the large size and rich information of personal genomic data, it is much more difficult to protect the genomic privacy of an individual than other sensitive information (such as social security numbers and names) that can be securely protected by encryption \cite{Malin04}. Therefore, it is highly required to develop efficient and fast methods for protecting genomic privacy while utilizing the genomic information for specifically designated purposes, such as medical diagnosis and new drug discovery.

Existing approaches for genomic privacy protection can be roughly divided into three categories: (1) cryptology-based methods \cite{Kantarcioglu08,Goodrich09}; (2) data de-identification methods \cite{Malin04,Malin00} and (3) data augmentation methods \cite{Lin02,Malin05}. Cryptology-based methods do not disclose raw genomic data while supporting the genomic data mining. However, this kind of methods are not suitable for long-term genomic privacy protection because the cryptographic algorithms can be broken in a comparably shorter time than the personal genomic privacy protection requires \cite{Humbert13}. Besides, they offer no protection against re-identification \cite{Loukides10}. Data de-identification methods tend to remove or encrypt those genomic data-associated identifiers which are also personally specific and sensitive, such as social security numbers or names. Nevertheless, these methods cannot guarantee sufficient privacy protection and are not able to deal with the re-identification problems \cite{Malin05b}. Data augmentation methods achieve the goal of privacy protection by generalizing or obfuscating DNA sequences, which can make each record indistinguishable from each other. With this kind of methods, the privacy of genomic data can be well protected at the expense of limited loss of data utility.

Among the aforementioned methods, a DNA sequence obfuscation method called DNA lattice anonymization (DNALA) \cite{Malin05} is one of the state-of-the-art approaches. DNALA is based on the famous $k$-anonymity principle \cite{Sweeney02} which uses a generalized sequence to represent $k$ aligned DNA sequences after sequence alignment and clustering. In this way, individual sequences within a cluster will not be distinguished. This method can efficiently protect the personal genomic privacy; however, it uses a low-accuracy clustering algorithm called CLUSTALW \cite{Thompson94} and a time-consuming sequence alignment technique. Later, Li. et al. \cite{Li07} proposed a stochastic hill-climbing method to improve the clustering algorithm for better performance. Recently, Li et al. \cite{Li12c} further reduced the information loss for genomic privacy protection by proposing a maximum-weight matching (MWM) based algorithm. However, these methods are still inefficient and with low accuracy.

To address these problems, this paper proposes a sequence-similarity based obfuscation method, namely IterMegaBLAST, for protecting personal genomic privacy. Unlike previous methods \cite{Malin05,Li07,Li12c}, which use CLUSTALW as the clustering algorithm, IterMegaBLAST uses MegaBLAST \cite{Zhang00} for both sequence alignment and clustering. MegaBLAST is a sequence alignment search algorithm which finds highly-similar sequences to the query one. Specifically, given a dataset, we iteratively use MegaBLAST to find homologs within the dataset for randomly selected query sequences. Then, the query sequences and the corresponding homologs are subsequently formed as clusters for further sequence obfuscation. Our results also demonstrate that IterMegaBLAST is much faster and more accurate than the existing state-of-the-art methods under the same degree of privacy protection.

\vspace{-0.2cm}
\section{Method}\label{sec:method}
\subsection{Problem Statement}
\vspace{-0.2cm}
\label{sec:problem}

Given a dataset of DNA sequences, our objective is to protect the individual-specific genomic information from identification and/or re-identification\footnote{Re-identification means matching the anonymized personal data with its original information or owner.} as much as possible while the loss of information affecting the data utility is as little as possible. In other words, the genomic privacy is enhanced at the expense of data precision reduction. One of the effective ways is to obfuscate the different information within a cluster of DNA sequences with high sequence similarity. In this way, the individual-specific privacy information can be preserved while the loss of information is the minimum. Due to their special properties, DNA sequences can not be clustered if without sequence alignment. Therefore, the procedures for a obfuscation method for genomic privacy protection generally include two steps: (1) sequence alignment and clustering; and (2) obfuscation (or anonymization).

\subsection{MegaBLAST for Sequence Alignment and Clustering}\label{sec:mega}
\vspace{-0.2cm}
MegaBLAST is a DNA sequence alignment search tool which uses a greedy algorithm \cite{Zhang00} to find those highly-similar sequences to the query one. MegaBLAST is optimized to find near identities and can provide functions of both sequence alignment and clustering. Compared to the traditional BLAST algorithm \cite{Altschul97}, MegaBLAST runs 10 times faster and is particularly efficient to handle much longer DNA sequences.\footnote{http://www.ncbi.nlm.nih.gov/staff/tao/URLAPI/new/node81.html} 

Therefore, MegaBLAST is very suitable for our case due to the following reasons: (1) the genomic data (i.e., DNA sequences) concerned should be aligned and clustered before obfuscation methods are used; (2) in practical situations, a fast sequence alignment and clustering tool is highly required to deal with a tremendous number of DNA sequences; (3) usually genomic privacy protection should be imposed on datasets of DNA sequences within the same species, which are often with high sequence similarity and MegaBLAST specifically excels in handling highly-similar sequence alignment.

Because MegaBLAST can find a list of homologs\footnote{A homolog is a sequence from a searching database which shares a high sequence similarity with the query one.} to the query sequence, we can select a certain number (i.e., the $k$ defined in Section~\ref{sec:k-anony}) of the top homologs together with the query sequence to form a cluster. Later, obfuscation methods are imposed on each cluster for genomic privacy protection.

\subsection{$k$-Anonymity}\label{sec:k-anony}
\vspace{-0.2cm}
The $k$-anonymity \cite{Sweeney02} was initially proposed to tackle a problem of how to make the individual data-owners indistinguishable while their data are publicly released and remain practically useful. The value $k$ refers to the number of individuals (or samples) within a cluster. In other words, the data are originally entity-specific and well-organized which are represented by some semantic categories (or attributes) consisting of a set of values. To prevent the data owners from being re-identified, a typical $k$-anonymity based method uses {\em generalization}. Generalization methods are based on a linear and unambiguous generalization hierarchy \cite{Malin05} where the value at the higher level (ancestor) is less-specific than that at the lower-level (child). They replace the value of each individual by a higher-level value via the generalization hierarchy rule. For example, we can use `California' to replace `Los Angeles' and `San Diego', and use `USA' to replace `California' and `New York'. In this way, a released data set processed by a $k$-anonymity method can guarantee that an individual's record within this data set cannot be distinguished from at least $(k-1)$ other individuals. In other words, the probability of re-identifying an individual based on the data set is no more than $1/k$. Obviously, a larger $k$ will provide better privacy protection. Besides generalization, suppression \cite{Kisilevich10} is another way to realize the $k$-anonymity.

\subsection{Sequence Obfuscation}\label{sec:obfuscation}
\begin{figure}[!t]
  \centerline{\epsfig{figure=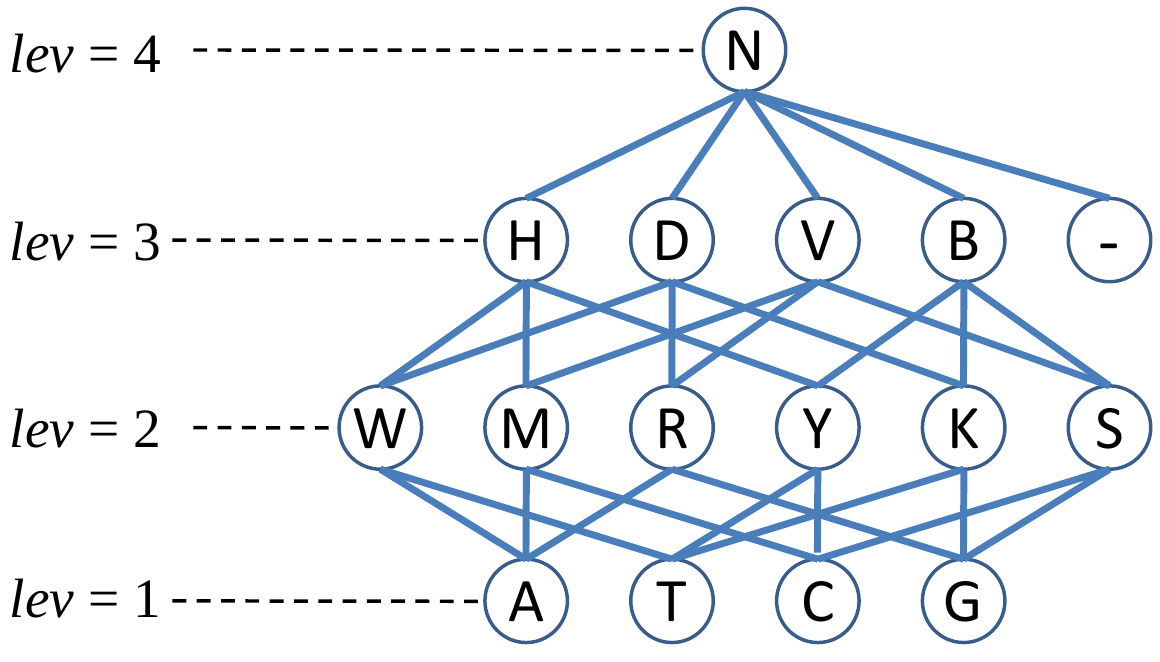,width=8.25cm}}
  \vspace{-0.1cm}
 \caption{\label{fig:code_hierarchy} The generalization hierarchy proposed in \cite{Malin05} for sequence/nucleotide obfuscation. Note that $lev$ is the level of corresponding nucleotides and the symbol `-' represents the gap.}
 \vspace{-0.1cm}
\end{figure}

\vspace{-0.2cm}
In this paper, a method proposed in \cite{Malin05} is used for sequence obfuscation. This method used a generalization hierarchy based on the IUPAC nucleotide representation code \cite{Comm70}. Generally speaking, the basic four nucleotides (\textsf{A}, \textsf{T}, \textsf{C} and \textsf{G}) act as the elements in the 1-st level of the generalization hierarchy; in the 2-nd level, six letters (\textsf{R, W, M, K, S} and \textsf{Y}) are used to represent the six different combinations of any two nucleotides in the 1-st level; letters (\textsf{D, V, H} and \textsf{B} as well as the gap) in the 3-rd level represent the combinations of any three nucleotides plus the gap; and we use the letter \textsf{N} in the 4-th level to represent all the possible situations. Details of the generalization hierarchy is shown in Fig.~\ref{fig:code_hierarchy}.

Specifically, given two nucleotides $q_{l}^{i}$ and $q_{l}^{j}$ in the $l$-th position of the $i$-th and the $j$-th aligned DNA sequences ${\mathds Q}_{(i)}$ and ${\mathds Q}_{(j)}$, respectively, their obfuscation (nucleotide) code is represented as $g(q_{l}^{i}, q_{l}^{j})$. For example, given two aligned nucleotide sequence segments \mbox{\textsf{CCTGTAAA}} and \mbox{\textsf{CA-GTRAA}}, according to the rule in Fig.~\ref{fig:code_hierarchy}, their obfuscation sequence is \mbox{\textsf{CMNGTRAA}}. To measure the information loss after sequence obfuscation, a distance measurement was proposed in \cite{Malin05}. The distance between $q_{l}^{i}$ and $q_{l}^{j}$ after nucleotide obfuscation is defined as:
\begin{equation}\label{eq:lev}
dist(q_{l}^{i}, q_{l}^{j}) = 2lev(g(q_{l}^{i}, q_{l}^{j})) - lev(q_{l}^{i}) - lev(q_{l}^{j}),
\end{equation}
where $lev(\cdot)$ is the level of nucleotides. Based on Eq.~\ref{eq:lev}, the distance between two aligned sequences (suppose the length of both sequences is $L$) can be defined as the sum of distances of all the nucleotides at the same positions, i.e.,
\begin{equation}\label{eq:alldist}
d({\mathds Q}_{(i)}, {\mathds Q}_{(j)}) =  \sum_{l=1}^{L} dist(q_{l}^{i}, q_{l}^{j}).
\end{equation}

Using the two sequences \mbox{\textsf{CCTGTAAA}} and \mbox{\textsf{CA-GTRAA}}, according to Eq.~\ref{eq:alldist}, we obtain the sequence distance is $d = 0 + 2 + 4 + 0 + 0 + 1 + 0 + 0 = 7$. In our experiments, we use the distance to measure the degree of information loss after sequence obfuscation. Definitely, the shorter the distance is, the less the information loss incurs after sequence obfuscation.

\subsection{IterMegaBLAST for Genomic Privacy Protection}\label{sec:itermega}
\vspace{-0.2cm}

Given a dataset of DNA sequences, the procedures for our method can be summarized in Algorithm~\ref{alg:mega}. In Algorithm~\ref{alg:mega}, $\floor*{x}$ means taking the largest integer less than or equal to $x$; $\cup$ and $\setminus$ are the set union and set difference, respectively; MegaBLAST$({\mathds Q}_{(t)}, {\mathcal S}_{(t)})$ means using ${\mathds Q}_{(t)}$ as the query sequence and ${\mathcal S}_{(t)}$ as the searching database to do the MegaBLAST search. Similar to other studies \cite{Li12c}, we set $k =2$ in our experiments. Note when the number of a dataset is odd, we need to use MegaBLAST to align the last three sequences. After sequence alignment, we obtain the obfuscated sequence for the query sequence and the top homolog. Then we do the second obfuscation on the second top homolog and the obfuscated sequence previously obtained.

For ease of reference, we name our method as IterMegaBLAST.
%Note that for different datasets, it is probably necessary to adjust the parameters (e.g., E, G and W) for MegaBLAST because (1) sequence lengths in different datasets may vary significantly and (2) it is necessary to strike a balance between achieving better sequence alignment results and imposing large gap penalties for shorter distances. In our case, we used E =5, G = 500 and W = 200.
\begin{algorithm}
%\SetKwInOut{Input}{Input}
%\SetKwInOut{Output}{Output}
%\KwData{this text}
%\KwResult{how to write algorithm with \LaTeX2e }
\caption{\label{alg:mega}The algorithm for IterMegaBLAST}
\textbf{Input:} A dataset ${\mathcal D} = \{{\mathds P}_{i}\}_{i=1}^{N}$ of $N$ DNA sequences.\\
\textbf{Output:} A set of obfuscated sequences ${\mathcal G}$ and a set of distances ${\bf d}$ between sequences and their obfuscated sequences.
\hrule
\begin{algorithmic}[1]
%\Statex \noindent \textbf{Input:} A dataset of $N$ DNA sequences, and parameters for MegaBLAST.
%\Statex \noindent \textbf{Output:} Obfuscated sequences and distances between sequences and their obfuscated sequences.
%\Statex \noindent \hrulefill
\State $t=0$;
\State ${\mathcal G} = {\bf d} = \varnothing$;
\State $R = [1,\ldots, N]$;
\State Set the initial dataset ${\mathcal D}_{0} = {\mathcal D}$;
\State Set the initial query and the homolog ${\mathds Q}_{(0)} = {\mathds H}_{(0)} =  \varnothing$;
\While{$t \leqslant (\floor*{N/2}-1)$}
\State $t \gets t + 1$;
\State ${\mathcal D}_{t} \gets {\mathcal D}_{t-1} \setminus ({\mathds Q}_{(t-1)} \cup {\mathds H}_{(t-1)})$;
\State \algparbox{Randomly select the $t$-th query sequence ${\mathds Q}_{(t)}$ from ${\mathcal D}_{t}$;}
\State Let the searching database be ${\mathcal S}_{(t)} = {\mathcal D}_{t} \setminus {\mathds Q}_{(t)}$;
\State \algparbox{Get the top homolog to the query ${\mathds Q}_{(t)}$ by MegaBLAST, i.e., ${\mathds H}_{(t)} = \mbox{MegaBLAST}({\mathds Q}_{(t)}, {\mathcal S}_{(t)})$;}
\State $R \gets R \setminus \mathop{\bigcup}_{i=1}^{N}{\{i: ({\mathds P}_{i} = {\mathds Q}_{(t)}) || ({\mathds P}_{i} = {\mathds H}_{(t)})\}}$;
\State \algparbox{Obtain the obfuscated sequence $g({\mathds Q}_{(t)}, {\mathds H}_{(t)})$ by the sequence obfuscation method stated in Section~\ref{sec:obfuscation};}
\State ${\mathcal G} \gets {\mathcal G} \cup g({\mathds Q}_{(t)}, {\mathds H}_{(t)})$;
\State \algparbox{Calculate the distance $d({\mathds Q}_{(t)}, {\mathds H}_{(t)})$ according to Eq.~\ref{eq:alldist};}
\State ${\bf d} \gets {\bf d} \cup d({\mathds Q}_{(t)}, {\mathds H}_{(t)})$;
%\State ${\mathcal D}_{i} \gets {\mathcal D}_{i} - {\mathds Q}_{(i)} - {\mathds H}_{(i)}$;

%Use MegaBLAST with ${\mathds Q}_{(i)}$ as the query and ${\mathcal D}_{(i)}$ as the database to produce a list of matched sequences (with high similarity) as well as alignment results between the query ${\mathds Q}_{(i)}$ and the matched ones;
%\State Select the top matched sequence (or top homolog) together with the query sequence ${\mathds Q}_{(i)}$ to form a cluster ($k=2$);
%\State Use the sequence obfuscation method to obtain the obfuscated sequence for the cluster in Step 7 and calculate the distance;
%\State Remove the two sequences within the cluster in Step 7 and the rest $(N-2*i)$ sequences as a new dataset;
%%\State Repeat Steps 4-9 until all sequences are exhausted;
\EndWhile
\State $t \gets \floor*{N/2}$;
\If{$N$ is odd}
\State $g({\mathds Q}_{(t)}, {\mathds H}_{(t)}) = g(g({\mathds P}_{R(1)}, {\mathds P}_{R(2)}), {\mathds P}_{R(3)})$;
\Else 
\State $g({\mathds Q}_{(t)}, {\mathds H}_{(t)}) = g({\mathds P}_{R(1)}, {\mathds P}_{R(2)})$;
\EndIf
%\State Repeat Steps 12-14;
\State ${\mathcal G} \gets {\mathcal G} \cup g({\mathds Q}_{(t)}, {\mathds H}_{(t)})$;
\State \algparbox{Calculate the distance $d({\mathds Q}_{(t)}, {\mathds H}_{(t)})$ according to Eq.~\ref{eq:alldist};}
\State ${\bf d} \gets {\bf d} \cup d({\mathds Q}_{(t)}, {\mathds H}_{(t)})$;
%\State Treat the final two or three sequences as a cluster. Repeat Steps 3-9 to obtain the obfuscation sequence and calculate the distance.
\end{algorithmic}
\end{algorithm}

%\begin{enumerate}
%  \item Given a dataset of $N$ DNA sequences, randomly select a DNA sequence as the $i$-th ($1 \leqslant i \leqslant \floor*{N/2}$)\footnote{$\floor*{x}$ means taking the largest integer less than or equal to $x$.} query sequence ${\mathds Q}_{(i)}$;
%  \item  The rest $(N-2*i+1)$ DNA sequences form the searching database ${\mathcal D}_{(i)}$;
%  \item Use MegaBLAST with ${\mathds Q}_{(i)}$ as the query and ${\mathcal D}_{(i)}$ as the database to produce a list of matched sequences (with high similarity) as well as alignment results between the query ${\mathds Q}_{(1)}$ and the matched ones;
%  \item Select the top matched sequence (or top homolog) together with the query sequence to form a cluster ($k=2$);
%  \item Use the sequence obfuscation method to obtain the obfuscated sequence for the cluster in Step 4;
%  \item Remove the two sequences within the cluster in Step 4 and the rest $(N-2*i)$ sequences as a new dataset;
%  \item Repeat Steps 1-6 until all sequences are exhausted;
%  \item In case the number of sequences is odd, treat the final three sequences as a cluster. Again use MegaBLAST to align these three sequences and obtain the obfuscation sequence of these three by the sequence obfuscation method.
%\end{enumerate}

\vspace{-0.2cm}
\section{Results and Discussions}\label{sec:results}
\vspace{-0.1cm}
\subsection{Datasets}
\vspace{-0.2cm}

Two datasets (Dataset I and Dataset II) \cite{Li12c} were used to evaluate the performance of IterMegaBLAST. Both datasets are human DNA sequences. Dataset I is a group of DNA sequences in the melanocortin gene promoter region while Dataset II is in the human mitochondrion control region. The numbers of sequences for these two datasets are 56 and 372, respectively. The average sequence length of Dataset I (6.6 kb) is much longer than that of Dataset II (0.5 kb).

The average distance between sequences and their obfuscated sequences, and the time complexity were used to measure the performance of different algorithms. Note that because all of the algorithms we compared in this paper are based on the $k$-anonymity, the degree of anonymity (or degree of privacy) \cite{Diaz02} should be the same when $k$ is the same. Therefore, we do not report the degree of privacy.

\vspace{-0.2cm}
\subsection{Performance of IterMegaBLAST Varying with respect to the Number of Sequences}
\vspace{-0.2cm}
\label{sec:performance}

%\begin{figure}[!t]
%  \centerline{\epsfig{figure=dist1_vs_num,width=8.25cm}}
%  %\vspace{-0.2cm}
% \caption{\label{fig:dist1_num} Average distances varying with respect to the number of DNA sequences for dataset I. The shorter the distance is, the less the information loss. DNALA is from \cite{Malin05}, while all of MWM , Hybrid and Online algorithms are from \cite{Li12c}.}
% %\vspace{-0.1cm}
%\end{figure}

\begin{figure}[!htbp]
\begin{minipage}[b]{1\linewidth}
  \centerline{\epsfig{figure=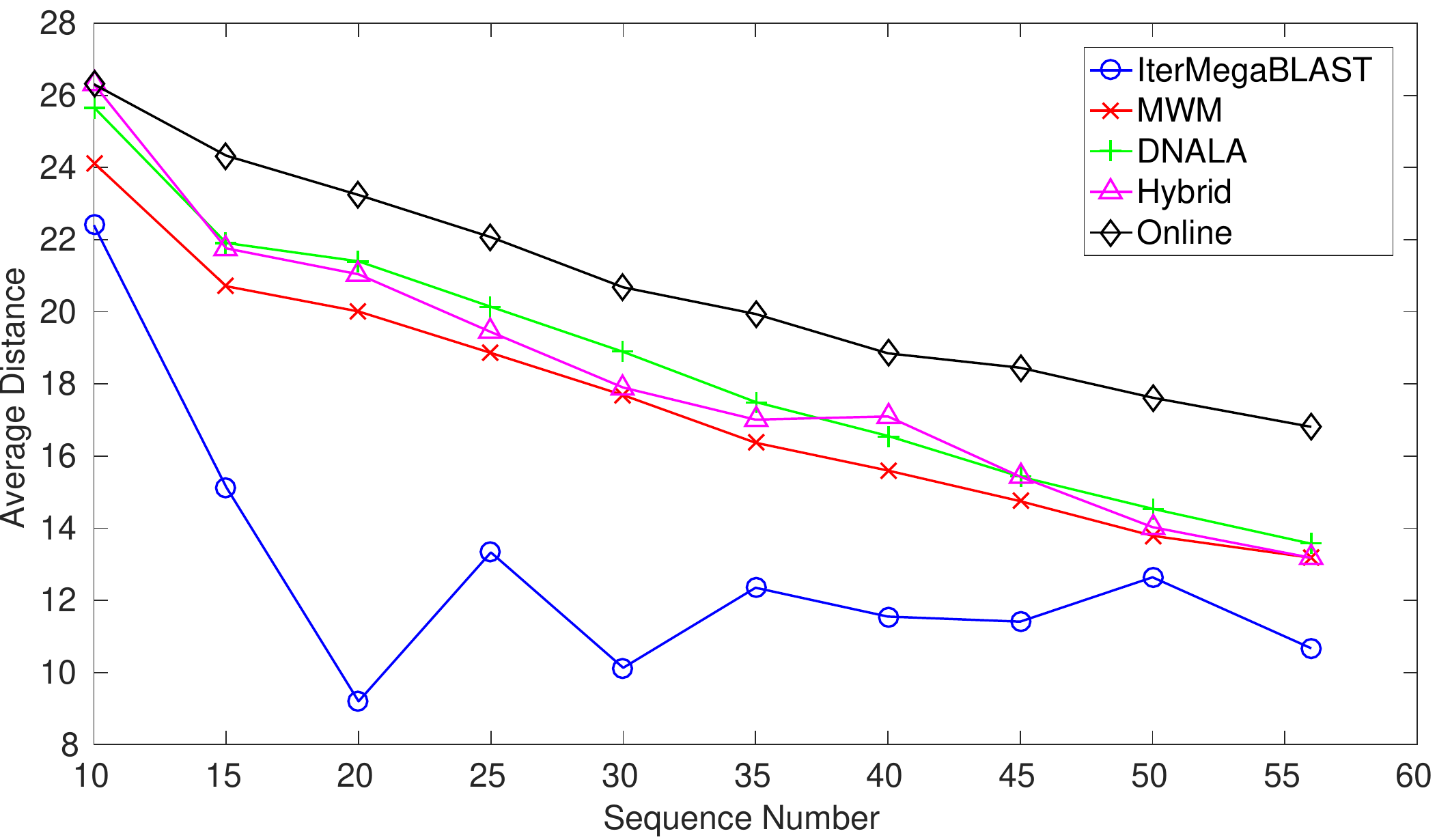,width=8.25cm}}
  \center{(a) Dataset I}
\end{minipage}
\begin{minipage}[b]{1\linewidth}
  \centerline{\epsfig{figure=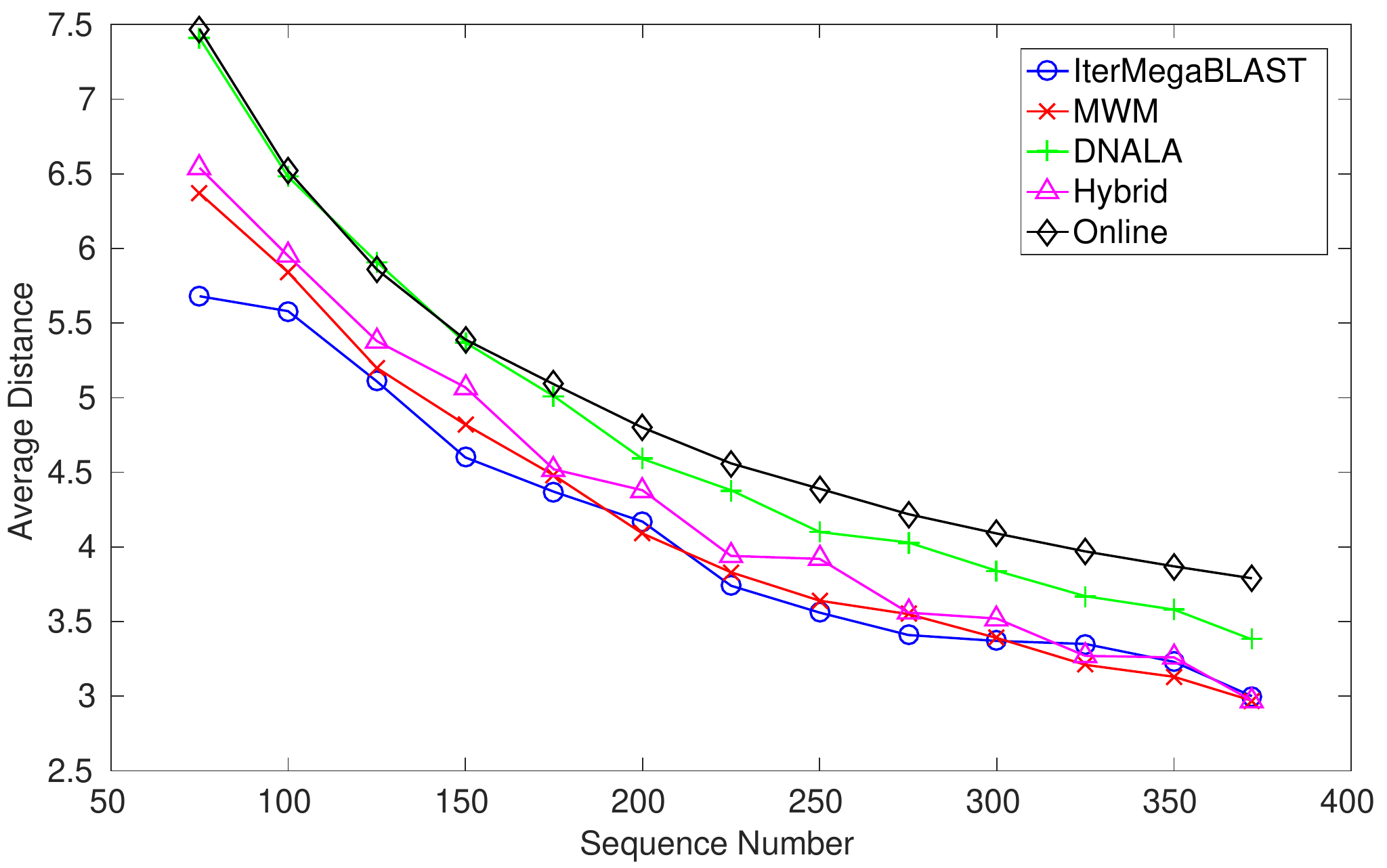,width=8.25cm}}
  \center{(b) Dataset II}
\end{minipage}

\caption{\label{fig:dist1_num}The average distances of IterMegaBLAST varying with respect to the number of DNA sequences for (a) Dataset I and (b) Dataset II. The shorter the distance is, the less the information loss. DNALA is from \cite{Malin05}, while all of MWM , Hybrid and Online algorithms are from \cite{Li12c}.}
\end{figure}

Fig.~\ref{fig:dist1_num} compares IterMegaBLAST against several state-of-the-art privacy-protection methods for both Dataset I and Dataset II when the number of DNA sequences gradually increase. DNALA \cite{Malin05} uses a multiple sequence alignment technique for sequence alignment and uses the CLUSTALW for clustering. All of MWM, Online and Hybrid use global pairwise sequence alignment, while for clustering, they use maximum weight matching \cite{Li12c}, an online algorithm \cite{Li12c} and hybrid of the former two algorithms. IterMegaBLAST uses an iterative MegaBLAST for both sequence alignment and clustering. The performance is measured by the average distances between sequences and their obfuscated sequences. The shorter the distance is, the less the information loss. Because the query DNA sequences for IterMegaBLAST are randomly selected, the performance of IterMegaBLAST may vary a bit even when the same DNA sequences are used. To reduce the bias, we performed IterMegaBLAST ten times for each case (number of sequences). For ease of presentation, only the average performance is shown.

As can be seen from Fig.~\ref{fig:dist1_num}(a), IterMegaBLAST significantly outperforms all of the state-of-the-art methods in all cases when the number of sequences increases from 10 to 56. While the average distances of all of MWM, DNALA, Hybrid and Online are strictly monotonically decreasing with the number of sequences, this is not the case for IterMegaBLAST, which achieves its best performance when the number of sequences is 20. It is noted that because all of these five methods are based on $k$-anonymity (i.e., $k =2$), the degree of anonymity \cite{Diaz02}, which is to measure the degree of how well the privacy is protected, should be the same. Therefore, experimental results suggest that under the same degree of anonymity, IterMegaBLAST can maintain the least information loss for data utility among all the genomic privacy-protection methods. The results also suggest that sequence similarity based methods (i.e., IterMegaBLAST) can provide sufficient privacy protection for genomic data (particularly long DNA sequences) while the information loss maintains at a low level.

Similar conclusions can be drawn from Fig.~\ref{fig:dist1_num}(b) except that IterMegaBLAST may be only comparable to (if not better than) MWM, particularly when the number of sequences is larger than 300. Except MWM, IterMegaBLAST performs better than DNALA and Online for all the ranges of sequence numbers, and outperforms the Hybrid algorithm for all cases except when the number of sequences is around 325. This is probably because the lengths of DNA sequences are vary short (average 0.5kb) and MegaBLAST is better able to handle long DNA sequences. Moreover, we would like to emphasize that the number of non-standard nucleotides (e.g., \textsf{N}) in the sequences of Dataset II is much larger than that of Dataset I, which contributes to more information loss whereas MegaBLAST treats them with equal weights as those standard nucleotides. On the other hand, MWM directly uses the minimum distance as the criteria to cluster the sequences.

\vspace{-0.2cm}
\subsection{Comparing with State-of-the-Art Predictors}
\vspace{-0.2cm}
\begin{table}[!t]
\footnotesize
\caption{\label{tb:compare1} Comparing IterMegaBLAST with state-of-the-art genomic privacy-protection methods. $m \pm n$ denotes (mean)$\pm$(standard deviation). The performance is measured by the average distance between DNA sequences and their obfuscated sequences. The shorter the distance is, the less the information loss. MSA: multiple sequence alignment; PSA: pairwise sequence alignment.}
\vspace{-0.3cm}
\begin{center}
{\begin{tabular}{|c|c|c|c|} \hline
 Dataset         & Method    & MSA    & PSA \\ \hline
 
\multirow{5}{*}{I}     & DNALA \cite{Malin05}     & 13.79  & 13.57 \\ \cline{2-4}
                 & MWM \cite{Li12c}       & 13.39  & 13.18 \\ \cline{2-4}
                 & Online \cite{Li12c}    & 16.93  & 16.81 \\ \cline{2-4}
                 & Stochastic hill-climbing \cite{Li07}     & 13.39  & 13.18 \\ \cline{2-4}
                 & IterMegaBLAST     & ---  & {\bf 10.67 $\pm$ 1.07} \\ \hline
\multirow{5}{*}{II}     & DNALA \cite{Malin05}     & 3.33  & 3.35 \\ \cline{2-4}
                 & MWM \cite{Li12c}       & 2.99  & {\bf 2.98} \\ \cline{2-4}
                 & Online \cite{Li12c}    & 3.79  & 3.80 \\ \cline{2-4}
                 & Stochastic hill-climbing \cite{Li07}     & 3.13  & 3.11 \\ \cline{2-4}
                 & IterMegaBLAST     & ---  & {\bf 3.00 $\pm$ 0.10} \\ \hline
\end{tabular}}
\end{center}
\vspace{-0.2cm}
\end{table}
%Table~\ref{tb:compare2} compares the performance of RP-SVM against several state-of-the-art multi-label predictors on the plant dataset. All of the predictors use the information of GO terms as features. From the classification perspective, Plant-mPLoc \cite{Chou10b} uses an ensemble OET-KNN (optimized evidence-theoretic K-nearest neighbors) classifier; iLoc-Plant \cite{Chou11b} uses a multi-label KNN classifier; mGOASVM \cite{Wan12b} uses a multi-label SVM classifier.
To further demonstrate the superiority of IterMegaBLAST, Table~\ref{tb:compare1} compares the performance of IterMegaBLAST against several state-of-the-art privacy-protection methods. Another algorithm called stochastic hill-climbing \cite{Li07} is added to compare with IterMegaBLAST. Moreover, DNALA, MWM, Online and Stochastic hill-climbing are capable of performing multiple sequence alignment (MSA) and pairwise sequence alignment (PSA).\footnote{We do not report the performance of the hybrid algorithm here because \cite{Li12c} does not provide the related results.} Note that it is also possible to do multiple sequence alignment for IterMegaBLAST and we will do it in our future research.

As can be seen from Table~\ref{tb:compare1}, for Dataset I, IterMegaBLAST remarkably outperforms all of the four state-of-the-art methods, no matter they use MSA or PSA; while for Dataset II, IterMegaBLAST performs better than DNALA, Online and stochastic hill-climbing, but its performance is comparable to (if not better than) that of MWM.

Table~\ref{tb:compare2} compares the computational time of IterMegaBLAST against MWM equipped with either PSA or MSA.\footnote{Note that the results of computational time for MWM are obtained from \cite{Li12c}. The configuration of our local computer may be different from that in \cite{Li12c}. However, we give credits to them by using only one core of our computer with a common configuration.} Since MWM performs the best among the four aforementioned methods, we only report the computational time of MWM here. As can be seen, IterMegaBLAST performs impressively faster than MWM + PSA and MWM + MSA for both datasets. The reason is that IterMegaBLAST only needs to use MegaBLAST for $\floor*{N/2}$ times and each time the number of sequences in the searching database will decrease. As we have mentioned, MegaBLAST performs 10 times faster than traditional BLAST, whereas MWM has to obtain all the pair-wise distances for all sequences. Interestingly, the computational time of IterMegaBLAST for Dataset II is much longer than that for Dataset I. This is because the number of sequences in Dataset II is much larger, causing a significantly larger number of MegaBLAST invocations for Dataset II. Moreover, MegaBLAST is more capable of handling long sequences like Dataset I, which also explains why the time advantage of IterMegaBLAST over MWM is more obvious for Dataset I than that for Dataset II.
\begin{table}[!t]
\footnotesize
\caption{\label{tb:compare2} Comparing the computational time of IterMegaBLAST with that of state-of-the-art genomic privacy-protection methods. MSA: multiple sequence alignment; PSA: pairwise sequence alignment.}
\vspace{-0.3cm}
\begin{center}
{\begin{tabular}{|c|c|c|c|} \hline
 Dataset         & Method    & Time (seconds) \\ \hline
\multirow{3}{*}{I}     & MWM + MSA \cite{Li12c}   & $>9000$ \\ \cline{2-3}
                 & MWM + PSA \cite{Li12c}           & $>7000$ \\ \cline{2-3}
                 & IterMegaBLAST                    & {\bf 112} \\ \hline
\multirow{3}{*}{II}     & MWM + MSA \cite{Li12c}  & $>2000$ \\ \cline{2-3}
                 & MWM + PSA \cite{Li12c}           & $>2000$ \\ \cline{2-3}
                 & IterMegaBLAST                    & {\bf 384} \\ \hline
\end{tabular}}
\end{center}
\vspace{-0.2cm}
\end{table}

\vspace{-0.2cm}
\section{CONCLUSIONS}\label{end}
\vspace{-0.2cm}

This paper proposes an accurate and efficient approach, namely IterMegaBLAST, which leverages sequence similarity and information obfuscation for genomic privacy protection. Given a dataset of DNA sequences, we formed clusters by iteratively selecting query sequences and finding their top homologs by MegaBLAST. Subsequently, the aligned sequences in each cluster were obfuscated by replacing the different nucleotides with their lowest common ancestors via a DNA generalization lattice scheme. It was found that IterMegaBLAST performs much better than existing genomic privacy-preserving methods with less information loss and higher efficiency under the same degree of genomic privacy protection.

\bibliographystyle{IEEEbib}

{ \bibliography{bioinform,mypapers,my_refs}}
\end{document}